\documentclass[aps,prl,reprint,groupedaddress]{revtex4-1}

\usepackage{graphicx}
\usepackage{dcolumn}
\usepackage{rotating}
\usepackage{amssymb}
\usepackage{mathptmx}
\usepackage{amsfonts}
\usepackage{amsmath}
\usepackage{bm}
\begin{document}

\title{Observation of Majorana Quasiparticles Surface States in Superfluid ${^3}$He-B by Heat Capacity Measurements}

\author{Yu.M. Bunkov$^{1}$, R.R. Gazizulin$^{1,2}$}

\affiliation{$^{1}$CNRS, Inst. NEEL, F-38042, Grenoble, France\\Univ. Grenoble Alpes, Inst. NEEL, F-38042, Grenoble, France\\
$^{2}$Kazan Federal University, 420008, Kazan, Russia}

\date{\today}

\begin{abstract}
We report about direct measurements of heat capacity of Majorana quasiparticles in superfluid ${^3}$He-B which appear near the surface of the experimental bolometer on the coherence length ${\xi}$. Two bolometers with different surface-to-volume ratios were used which allows us to have different calibrated contributions from Majorana quasiparticles to the ${^3}$He heat capacity. Estimations of possible impact of ${^3}$He layers adsorbed on the walls of the bolometer have been done.
\end{abstract}



\maketitle

The main property of Majorana fermions predicted in 1937 \cite{Majorana} is that particles are identical to their own antiparticles. While no elementary particle is established to be Majorana fermion there are quasiparticles in some special classes of condensed-matter systems which have Majorana properties \cite{RevModPhys, Nature1, Nature2}. Particularly, these quasiparticles appear at the boundaries of superfluid ${^3}$He-B \cite{Volovik}. 

At a critical temperature $T_{c}$, varying between 0.93 and 2.49 mK as a function of pressure, ${^3}$He passes a phase transition to its superfluid states, unique in their richness. These states show a lot of analogies to superconductors which can be regarded as the superfluid state of electron liquid, but also some important differences which makes ${^3}$He even more fascinating. The key factor of superfluidity is the formation of Cooper pairs, i.e. pairs of ${^3}$He quasiparticles, which form the ground state of the system. The dispersion relation for the elementary excitations, i.e. broken Cooper pairs (the Bogolyubov quasiparticles (QPs)), above this state gets:

\begin{equation}
  E=\sqrt{\eta ^{2}+\Delta ^{2}},
   \label{dispersion}
\end{equation}
where $\Delta$ is the temperature dependent gap parameter and $\eta$ is the kinetic energy of the excitation. The striking feature of this dispersion relation is the fact that no excitations with $E<\Delta$ exist. 

In case of B-phase of ${^3}$He the gap is isotropic. As was shown in \cite{JETP1, JETP2}, this phase supports the existence of Majorana QPs, which follow from the particle-hole symmetry of the Bogolyubov QPs, $\gamma _{E}^{+}=\gamma _{-E}$. It is straightforward to see that at zero energy $\gamma _{0}^{+}=\gamma _{0}$, which informs us that the zero-energy quasi-particles are the same as the quasi-holes and are therefore Majorana QPs. These states appear near the walls where the energy gap is suppressed exactly to zero $\Delta = 0$ over a distance corresponding to the superfluid coherence length $\xi$ (Fig. 1).

 \begin{figure}[htt]
 \includegraphics[width=0.55\textwidth]{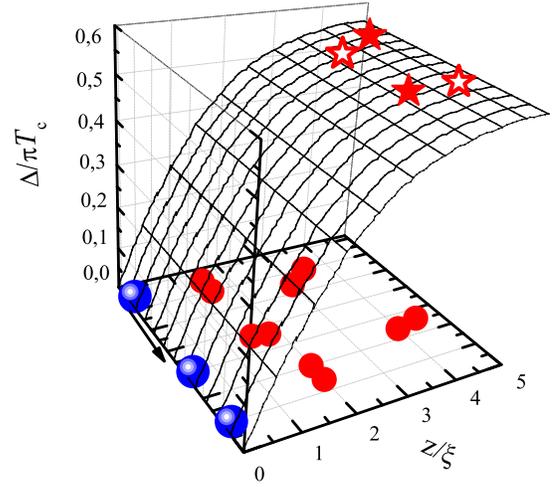}
 \caption{The energy gap of superfluid ${^3}$He-B as a function of distance from the wall surface in coherence length units. Bogolyubov QPs (shown by stars) above the gap can move in the bulk ${^3}$He. Cooper pairs (shown by red balls) form the ground state of system. Majorana QPs are located near the surface on the coherence length distance where the superfluid gap is exactly zero, and they can move along the surface due to non-zero kinetic energy contributing to the ${^3}$He heat capacity.}
 \label{gap}
 \end{figure} 

The presence of Majorana surface states in ${^3}$He-B can be probed through anomalous transverse sound attenuation \cite{sound1, sound2, sound3, sound4} and surface heat capacity measurements \cite{Mizushima, Halperin, Bunkov1, arxiv, Bunkov2}. A direct consequence of the presence of the gap in energy spectrum is the exponentially decreasing of heat capacity of Bogolyubov QPs:
\begin{equation}
  C_{bulk} \sim V \, P_F^2\,  (\frac{\Delta}{kT})^{3/2} \exp \;( - \frac{\Delta}{kT} )
   \label{Cbulk}
\end{equation}
where $P_F$ is the Fermi momentum, $\Delta$ is the superfluid gap $ \simeq 2kT_c$ and $V$ is the volume of the sample.

Zero energy gap for Majorana QPs leads to the power law dependence of heat capacity on temperature \cite{private}:
\begin{equation}
  C_{maj} \sim S \, \xi \, P_F^2\,  (\frac{\Delta}{kT})^{-2},
   \label{Cmaj}
\end{equation}
where $\xi$ is the $^3$He-B coherence length and $S$ is the surface area of the sample.

The ratio of these heat capacities, including the numerical factors,
reads:
\begin{equation}
\frac{C_{maj}}{C_{bulk}} = \frac{\pi^3}{8\sqrt2} \frac{ \xi S}{V} \,
(\frac{\Delta}{kT})^{-7/2}\exp \, ( \frac{\Delta}{kT} ).
   \label{Cratio}
\end{equation}

From Eq. (3) it can be seen that depending on surface-to-volume ratio Majorana QPs may have significant contribution to the ${^3}$He heat capacity by comparison with heat capacity of Bogolyubov QPs from some low enough temperature. Particularly, for used in our experiments bolometers with $S/V\sim$ 0.9 (1/mm) and 18.8 (1/mm) Majorana QPs should have 50$\%$ contirubution to the whole heat capacity at about 120 and 160 $\mu$K, respectively. Taking into account that dependence of the coherence length and superfluid gap on magnetic field and pressure is well tabulated the experimental investigations of such dependences is also the perspective task to distinguish heat capacity of Majorana QPs. The problem for magnetic filed dependence measurements is that in our case the experimental cell is located in the same field as the one used for the nuclear demagnetization. Due to this fact the magnetic field can not be changed without changing temperature. That is why the strategy was to go to optimal temperature and suppose that as the magnetic field is weak (magnetic energy is much weaker than superfluid gap) the results do not depend on it too strongly. The pressure dependence experiments will be published elsewhere. All results presented in this article were obtained at 0 bar. 

To measure heat capacity of superfluid ${^3}$He inside the bolometer it is necessary to release some known amount of energy and to measure the corresponding temperature response. For both purposes we used the Vibrating Wire Resonators (VWRs) which were superconducting NbTi wires fixed at both ends in the plane and of an approximately semicircular shape. The use of VWR as a thermometer has the advantage to give direct access to the properties of the liquid, without an intermediary liquid-solid boundary. The VWR techniques were first described in \cite{Lancaster}. It measures directly the density of quasiparticle excitations in superfluid ${^3}$He. As the density of Bogolyubov QPs follows the exponential law, its measurements allows a precise determination of the temperature by measuring the Full Width Half Maximum $W$ of the VWR
\begin{equation}
  W(T)\sim \alpha \exp (-\Delta /kT),
   \label{wire}
\end{equation}
where $\alpha \sim 10^5$ Hz is pre-factor which depends on geometry of the wire and pressure. Everywhere in the following analysis we used the Lancaster calibration \cite{LancCalib} for pre-factor $\alpha$ to obtain the temperature. 

For energy releasing inside the bolometer another VWR was used. This method of ${^3}$He heating, firstly proposed in \cite{heating}, is based on the mechanical friction of a wire. By driving it with an excitation current at its resonance frequency during a short time, a controllable amount of energy can be injected into the system. The amount of energy introduced electrically to the wire can be calculated by the integral of the electric power $E_{electric}=\int UIdt$. The electric power first transforms into kinetic/potential energy of the heater wire, and then dissipates to fluid via the velocity dependent frictional coupling. 

The bolometer consists of a cooper box filled with ${^3}$He. One VWR (measuring VWR) inside the cell is used as the thermometer while another VWR (heating VWR) inside the same cell is used as the energy source. The box has a tiny orifice at one of its sides which is used as the thermal link to the surrounding ${^3}$He, representing the heat sink. Due to the Kapitza resistance, the ${^3}$He is thermally decoupled from the cell walls, and the thermalisation only takes place through the orifice. After reaching the working temperature of about 100 $\mu$K the measuring VWR is switched on monitoring mode which allows to determine the temperature with good time resolution less than 1 s. Then heating pulse of about 70 ms is applied to the heating VWR which introduce the energy of about 100-1000 keV into the bolometer. After the heating pulse the temperature inside the cell rises and then goes back to its initial temperature by thermalisation via the orifice with some time constant. This temperature response is seen by measuring VWR by increasing of its resonance width. To make relation of the increasing of the VWR width $\Delta W$ after the heating pulse to the corresponding heat release $U$ let us define the calibration factor
\begin{equation}
  \sigma =\frac{\Delta W}{U}=\frac{1}{C(T)}\frac{dW(T)}{dT},
   \label{wire}
\end{equation}
where $C(T)$ is the heat capacity of ${^3}$He, $W(T)$ is obtained by eq. (5). 

Two bolometers were used in our experiments. The main difference between them is that they have unequal surface-to-volume ratios. In the first one (cell $A$) $S \sim$ 113 mm${^2}$ and $V \sim$ 130 mm${^3}$, while in the second (cell $B$) $S \sim$ 3216 mm${^2}$ and $V \sim$ 171 mm${^3}$. The surface in the cell $B$ was increased by adding the cooper slabs which have the thermal contact with the walls of the bolometer. Another important difference between cells concerns the diameters of measuring VWRs. In the cell $A$ we used the 4 $\mu$m VWR and in the cell $B$ - 10 $\mu$m. It was done due to following reason. It is known that the diameter of VWR defines the temperature range where the wire has the best sensitivity. Indeed, the less diameter of the wire the more dilute gas of quasiparticles (the lower temperature) it could sense. Following the eq. (4) we should expect $\it{a}$ $\it{priori}$ that in the cell $B$ the heat capacity of Majorana QPs may have the significant contributions at higher temperatures than in the cell $A$. Because of the crossover when the Majorana QPs heat capacity begins to deviate from the bulk Bogolyubov QPs heat capacity is very important temperature region to measure we installed the wire with bigger diameter into the cell $B$. 

The temperature dependences of calibration factors for both cells at 0 bar are shown in Fig. 2 (data for cell $A$ is taken from \cite{arxiv}). All necessary corrections, such as the finite response time of measuring VWR and the temperature dependence of the thermalisation time after heating pulse, have been done. The theoretical dependence of the calibration factor $\sigma \sim 1/\sqrt{T}$ taking into account only contribution of Bogolyubov QPs (eq. 2) is shown by dashed line (black for cell $A$ and blue for cell $B$). No fitting parameter was used to build this theoretical curves. The theoretical bulk calibration factor in the cell $B$ is less than in the cell $A$ because of the bigger diameter of the measuring wire (leads to the decreasing of the pre-factor in eq. (5)) and bigger volume of the cell (leads to the increasing of ${^3}$He amount inside the cell and, hence, to the increasing of the total heat capacity of ${^3}$He). 

 \begin{figure}[htt]
 \includegraphics[width=0.53\textwidth]{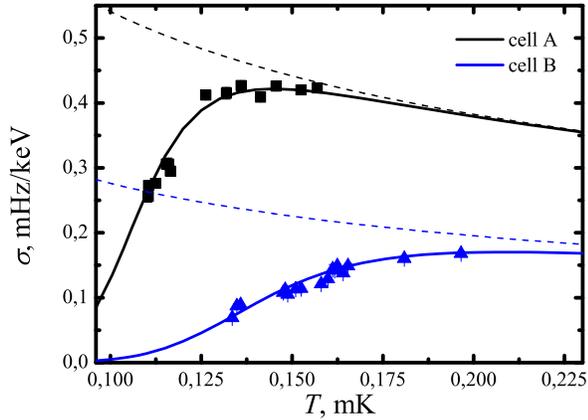}
 \caption{Calibration factors of measuring VWRs in two cells with different surface-to-volume ratio at 0 bar. Dashed lines correspond to the theoretical temperature dependence for Bogolyubov QPs only, while solid lines include additional contribution from Majorana QPs.}
 \label{sigma}
 \end{figure}
 
 It can be seen in Fig. 2 that the experimental data deviate from the dashed theoretical curve for both cells beginning from some temperature which is different for the cells. Eq. (6) shows that this drastic deviation can be explained either by temperature dependent losses of electrical power $U$ inside the heating wire or by some additional heat capacity. Indeed, the very similar experimental results in the cell $A$ were previously obtained during investigations of Dark Matter detector \cite{DarkMatter}. The authors explained the deviation of the calibration factor temperature dependence from the $1/\sqrt{T}$ law by energy losses inside the heating VWR. In our experiments in both cells we observed the clear saturation of the amplitude of heating VWR during its excitation which shows that the critical velocity for Cooper pair breaking is achieved. Consequently, the mechanical energy transfer to the quasiparticles is very effective and we should assign all the energy dissipated in heating VWR as contributing to the ${^3}$He. Moreover, attempts to apply the paradigm of intrinsic losses in heating VWR to interpret the results obtained in cell $B$ where the deviation begins at higher temperatures leads to the enormous parameter of internal friction processes of about 3 Hz which two orders of magnitude bigger than for cell $A$ with $W_{int} \sim $ 77 mHz. Despite of difference of the diameter of the wires used in the cells $A$ and $B$ for heating (13 and 10 $\mu$m, respectively) there are no clear reasons for that huge difference in internal friction processes. In addition, this theoretical curve does not fit adequately experimental data obtained in the cell $B$.
 
 Indeed, the temperature dependence of the calibration factors in both cells can be explained by the additional heat capacity of surface Majorana QPs (Eq. 3). The corresponding theoretical curves are shown by solid lines in Fig. 2 (black for cell $A$ and blue for cell $B$). We used the roughness of the cooper walls as a fitting parameter. It turned out that good agreement with experimental data when the roughness factor is equal to 10 for the both bolometers. Electron microscopy of the cell walls shows the high roughness so that this value of the factor seems to be reasonable. No any other fitting parameter was used. The only quantitative difference between theoretical curves for the two cells goes from the different $S/V$ values responsible for ratio between Majorana and Bogolyubov QPs heat capacities (Eq. (4)). It can be seen in Fig. 2 that these curves follow the experimental data points with excellent agreement. 
 
 The experimental data were recalculated to the corresponding heat capacities by using Eq. (6) and then normalised to the bulk Bogolyubov QPs heat capacity. These points are shown in Fig. 3, as well as the normalised heat capacities for Majorana QPs (dashed curved lines) and Bogolyubov QPs (dashed line equal to 1). As in Fig. 2, black lines correspond to the parameters of the cell $A$ and blue lines for those in the cell $B$. 
  
 \begin{figure}[htt]
 \includegraphics[width=0.53\textwidth]{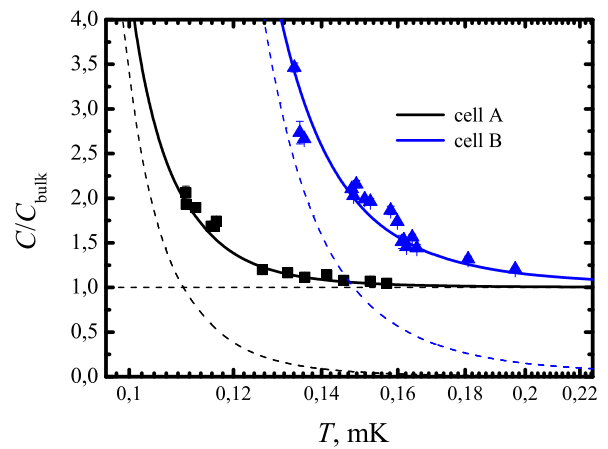}
 \caption{The heat capacity ratio $C/C_{bulk}$ calculated from data shown in Fig. 1. The line equal to 1 corresponds to Bogolyubov QPs. The black (cell $A$) and blue (cell $B$) curved dashed lines are that for the Majorana QPs, and the corresponding solid lines is for the heat capacity of both components.}
 \label{HeatCapacity}
 \end{figure}

It can be seen that in the both cells measured heat capacity follows to the sum of the Majorana and Bogolyubov QPs contributions. Moreover, in the cell $B$ at the temperature of about 130 $\mu$K the Majorana QPs contribution to the ${^3}$He heat capacity is four times bigger than that of the Bogolyubov QPs in full agreement with the estimations by Eq. (4). It allows us to conclude that we directly see the Majorana QPs in superfluid ${^3}$He-B.
 
 The question that could be stated now if there are any other possible sources of additional heat capacity besides the Majorana QPs. The fact that we measured the surface states and determined the strong influence of $S/V$ ratio leads to the assumption about possible influence of the ${^3}$He layers adsorbed on the walls of the cell. Indeed, in our experiments this influence was strongly suppressed because the surface had been covered by ${^4}$He during the condensation of the liquid ${^3}$He into the cell. One can suggest that there are some islands of adsorbed ${^3}$He remained in the cell which could give the additional heat capacity to the bulk ${^3}$He. But there are experimental facts that show that it is not the case.The precise measurements of the heat capacity of adsorbed ${^3}$He in the interesting for current experiments temperature range were done in \cite{heat_capacity}. It was found out that the heat capacity of adsorbed ${^3}$He has the magnetic nature. In this case of the ensemble of ${^3}$He nuclei \cite{Lounasmaa} the magnetic heat capacity decreases with increasing temperature as $C_{magn}\sim 1/T^{2}$. The dependence $C_{bulk}+C_{magn}$ is shown in Fig. 4 by red line. The amount of adsorbed ${^3}$He atoms was used as a fitting parameter. It should be mentioned that the fitting parameter obtained for the best fitting in Fig. 4 approximately correspond to fully coverage of the cell walls by ${^3}$He. While there are no reasons for so rough assumption because of the presence of ${^4}$He atoms on the walls it can be seen that the temperature behaviour is completely different from obtained in the experiments which has rather Majorana-like dependence shown by solid blue line. 
 
 \begin{figure}[htt]
 \includegraphics[width=0.53\textwidth]{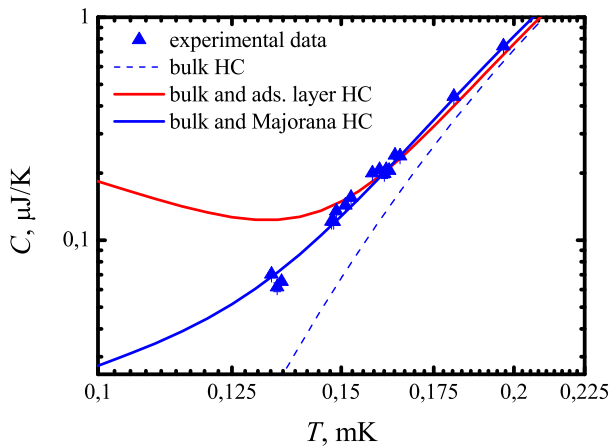}
 \caption{Heat capacity of superfluid ${^3}$He in dependence on temperature in the cell $B$. The dashed blue line depicts the exponential behaviour in the bulk region. The solid blue line includes the contributions of the Majorana-like dependence. The red line show the temperature dependence of heat capacity taking into account possible impact of ${^3}$He adsorbed layers. In the latter case the amount of the adsorbed atoms was used as a fitting parameter.}
 \label{HeatCapacity}
 \end{figure}

Another reason which allows us to exclude the impact of adsorbed ${^3}$He is that the temperature decay after the typical heating pulse in our experiments is well described by single exponential function while the presence of the adsorbed ${^3}$He leads to the two exponential behaviour of the decay after heating event \cite{heat_capacity, Elbs} due to the imperfect thermal contact between the solid and liquid ${^3}$He.
 
 Thus, we can make the conclusion that the existence of Majorana QPs is experimentally confirmed in the superfluid ${^3}$He-B by the direct measurements of its heat capacity in the cells with different surface-to-volume ratios where it has various contributions to the bulk exponential dependence on temperature.

\begin{acknowledgments}
We would like to thank G.R. Pickett, G.E. Volovik and C.B. Winkelmann for very useful discussions. This work was supported by the "Agence Nationale de la Recherche" (France) within MajoranaPRO project (ANR-13-BS04-0009-01). The work is performed according to the Russian Government Program of Competitive Growth of Kazan Federal University.
\end{acknowledgments}

\bibliography{basename of .bib file}

\end{document}